# Deep Learning for Prostate Pathology

Okyaz Eminaga [1,2,3,4], Yuri Tolkach*[5], Christian Kunder* [6], Mahmood Abbas*[7], Ryan Han[8], Rosalie Nolley [2], Axel Semjonow[9], Martin Boegemann[9], Sebastian Huss[10], Andreas Loening [11], Robert West [6], Geoffrey Sonn[2], Richard Fan[2], Olaf Bettendorf [7], James Brook [2] and Daniel Rubin [3,4]

*Equally contributed.

1) DeepMedicine.ai

2) Dept. of Urology, Stanford University School of Medicine

3) Dept. of Biomedical Data Science, Stanford University School of Medicine

4) Center for Artificial Intelligence in Medicine & Imaging, Stanford Medical School

5) Dept. of Pathology, Bonn University Hospital

6) Dept. of Pathology, Stanford University School of Medicine

7) Institute for Pathology and Cytology, Schuettorf, Germany

8) Dept. of Computer Science, Stanford University

9) Prostate Center, Dept. of Urology, University Hospital Muenster, Germany

10) Dept. of Pathology, Department of Urology, University Hospital Muenster, Germany

11) Dept. of Radiology, Stanford University School of Medicine

**Corresponding author:**

Okyaz Eminaga, M.D./Ph.D.

Department of Urology, Center for Artificial Intelligence in Medicine & Imaging (AIMI),

Laboratory of Quantitative Imaging and Artificial Intelligence (QIAI)

Stanford University School of Medicine

300 Pasteur Drive

Stanford CA 94305-5118

Tel: 650-725-5544

Fax: 650-723-0765

Email: okyaz.eminaga@stanford.edu

Stanford University, Stanford Medical School

# Abstract

The current study detects different morphologies related to prostate pathology using deep learning models; these models were evaluated on 2,121 hematoxylin and eosin (H&E) stain histology images captured using bright field microscopy, which spanned a variety of image qualities, origins (whole slide, tissue micro array, whole mount, Internet), scanning machines, timestamps, H&E staining protocols, and institutions. For case usage, these models were applied for the annotation tasks in clinician-oriented pathology reports for prostatectomy specimens. The true positive rate (TPR) for slides with prostate cancer was 99.7% by a false positive rate of 0.785%. The F1-scores of Gleason patterns reported in pathology reports ranged from 0.795 to 1.0 at the case level. TPR was 93.6% for the cribriform morphology and 72.6% for the ductal morphology. The correlation between the ground truth and the prediction for the relative tumor volume was 0.987 n. Our models cover the major components of prostate pathology and successfully accomplish the annotation tasks.

# Introduction

Prostate Cancer (PCa) is the most diagnosed cancer in men and one of the most prevalent cancer-related causes of death[1]. PCa is usually diagnosed via prostate needle biopsy and may result in patients undergoing radical prostatectomy (total removal of the prostate, seminal vesicles and surrounding tissues) upon histological confirmation[2]. Management of patients requires a reliable histopathological evaluation, including an initial determination of tumor extent and other cancer-related metrics (particularly grading). However, the limited human resources and the increase in the workload challenge the pathologists to maintain their evaluation performance during the clinical routine. Moreover, the prostatectomy specimens are processed in multiple embeddings (up to 50) and represent one of the specimens that have the most time-consuming evaluation process in the anatomical pathology.

The results from the histopathological evaluation are critical for decision-making and predicting the patient's outcome, making reproducibility and standardization of clinical importance [3]. While the histopathological evaluations of the prostatectomy specimens for the grading and staging of PCa are based on well-established guidelines[4,5], standardizing the pathological evaluation has proven to be a challenge due to factors such as the substantial interinstitutional differences in laboratory techniques and human reliability (intra-/interobserver variability)[5,6]. Pathologists are human beings whose mental capacities, visual perception, and responses are intuitively expected to vary at the individual level and are also influenced by socioeconomic conditions. Moreover, the histopathological evaluation generally represents a visual search and decoding task that depends on the human attentional capacity. It has been shown that human observers often differ in response to the same sensory stimuli for the same task, which explains the intra/interobserver variation in tumor grading [4,6,7]. Interestingly, advance information has been shown to enhances the visual search performance [8]. Moreover, providing prior information about the sample origin and location, in addition to the clinical data, has shown to improve the pathological evaluation [9,10]. Accordingly, we propose that providing prior information about the tumor extension and morphology can be helpful to guide a more precise histopathological evaluation. Recent advances in Artificial intelligence (AI) especially in computer vision has demonstrated its potentials for automated cancer detection and the tumor grading from histology images [11-17]. Deep learning (DL) is a board family of the machine learning methods within the AI domain. DL is considered as one of the state of the art algorithms in computer vision due to its remarkable performance in vision detection and segmentation tasks [18]. Most published works to date utilized publicly available "state-of-the art" neural network architectures like VGG16 [19], Inception V3 [20], ResNet [18], DenseNet [21] for tumor detection and grading problems. Several works have successfully shown the effectiveness of DL models in determining cancer lesions and performing tumor grading for PCa [11-17]. However, applying such DL model architectures to the cancer detection and grading task is hindered by the need for expensive computational resources and the absence of well-annotated development datasets. An ideal DL model for the medical domain would be trainable on a small or mid-size data set using affordable or existing infrastructure. Transfer learning is an approach to train a model on small/midsize datasets by reusing the pre-trained weights from a large dataset designed for a general image classification problem (e.g., ImageNet). However, the histology images are domain specific datasets that differ significantly from the dataset for the general image classification problems. Accordingly, optimizing the transfer learning remains challenging for the cancer detection problems from the histology images. Transfer learning further requires a complicated fine-tuning of the pretrained models like, for example, identifying the right layers to train the layer weights [22,23]. Thus, we chose to use PlexusNet, a customized deep learning architecture, that provides comparable results to "state-of-the-art" models for prostate cancer detection with limited resources[23]. For a real case usage, PlexusNet and other customized convolutional neural network architecture inspired by VGG were used for automated annotation to disentangle the annotation work from the pathologist's tasks. For that purpose, we utilized a framework based on cMDX© (Clinical Map Document based on XML) for the generation and management of clinician-oriented pathology reports already introduced by Eminaga et al [24]. The cMDX framework has been applied in clinical routine to reporting and analyses of PCa in prostatectomy specimens [25]. In this context, the topographical distribution of PCa foci and related pathologic findings can be evaluated using the cMDX documentation system [24,25]. However, our previous work was limited by the dependency on the pathologists who delineated the tumor extension using the cMDX Editor. Additionally, the annotation work and the documentation of Gleason patterns and pathological morphology (i.e. cribriform and ductal morphology) for each lesion remained time-consuming and are thus often times avoided by the pathologists. Given the restrictions of the previous works, this study will illustrate how AI can improve the existing framework utilized in clinical routine by taking over the annotation tasks and provide initial information for pathological evaluation.

## Results
### Prostate Cancer and related Findings

The interobserver annotation agreement between CK and OK/YT was acceptable by an average Cohen-Kappa score of 0.8385 (range: 0.7468 - 0.9284). The detection model for prostate cancer was validated internally and externally on 3 datasets representing the digitalized whole-mount (WM) slide images, the regular whole-slide (WS) images, and TMA. The comparison analyses to baseline methods (e.g. Inception V3 and MobileNet V2) related to prostate cancer detection are already handled by Eminaga et al [23]. Although these datasets were acquired using different types of scanners, PCa detection achieved AUC-ROCs range between 0.954 and 0.957 or Brier score range between 0.046 and 0.134 per slide for WM/WS images or at spot level for TMA (**Figure 1 and Table 1**). AUC-ROC reveals the classification performance at different thresholds; a higher AUC-ROC indicates a better classification accuracy, where a AUC-ROC of 1 represents the highest accuracy; the Brier score is used as a measure of the model "calibration"; the lower the brier score, the better the model is calibrated. A coefficient of determination ($R^2$) of 0.987 was measured for the correlation between relative tumor volumes of the ground truth (the tumor annotation made by CK for WM images of each case was considered as ground truth) and the predicted relative tumor volumes in 46 cases (**Supplement File 1)**. The paired t-test also showed no significant differences between the relative tumor volumes of the ground truth and the predicted relative tumor volumes (t-statistic: -0.499; P=0.619). The mean relative tumor volume was 9.95% for the ground truth and 11.0% for the predicted volume; the mean difference between both relative volumes was -1.08% (95% CI: -1.44 - 0.72). Our model correctly classified the slides in 99% (402/406) of the cases. The positive predictive value (PPV) was 99.2% and the negative predictive value (NPP) was 95.8%. The true positive rate (TPR) was 99.7%, while the false positive rate was 0.785% (**Supplement File 1**).

*[Removed due to copyright issue]*

*Figure 1 shows the general workflow for the detection part of the current study. For simplicity, we presented the results from prostate cancer and for the pathology features Gleason patterns 3 and 4 and HGPIN (High-grade intraepithelial neoplasia). The pathology reports do not routinely include HGPIN as the clinical benefits of HGPIN are limited on final pathology reports of*

prostatectomy specimens. Detailed results are provided in supplement file 1. The tumor burden was calculated by identifying the average number of pixels with tumor in relation to the pixel size on the mask patches generated from the annotation data in each data source. w/o: without; w: with; PCA: Prostate cancer; PPV: Positive predictive value; NPV: negative predictive value; TPR: true positive rate, TNR: true negative rate; CA: Classification accuracy; AUC: Area under curve of the receiver operating characteristic curve, FMI: Fowlkes-Mallows index. * a complete ground truth (annotation) for the tumor extent of the prostate cancer was available for 46 cases. TCGA: The Cancer Genomics Atlas;

Table 1: The slide-wise accuracies for prostate cancer detection on external datasets calculated on patch images per slide.

| Dataset | Scanner | Number of Images | AUC (95% CI) | Brier score (95% CI) |
| --- | --- | --- | --- | --- |
| Whole-mount slide images | A | 254 | 0.955 (0.953 - 0.956) | 0.046 (0.034 - 0.059) |
| Whole-slide images | B | 13 | 0.957 (0.939 - 0.975) | 0.116 (0.073 - 0.159) |

Table 2: The accuracies for Gleason pattern detection, cribriform and ductal morphology on cases that have whole-mount sections of the prostate. The frequency thresholds of the finding presence required for reporting at a case level were set to 0.1% for most findings.

| Finding | Total cases | Negative (%) | Positive (%) | Frequency threshold for positive patch images | Correctly detected by the pipeline | | CA | Recall | Precision | F1-score |
| --- | --- | --- | --- | --- | --- | --- | --- | --- | --- | --- |
| | | | | | TN (%) | TP (%) | | | | |
| Gleason Pattern 3 | 55 | 6 (10.9) | 49 (89.1) | 1 per 1000 patch images | 5 (83.3) | 49 (100) | 98.2 | 1.0 | 0.833 | 0.909 |
| Gleason Pattern 4 | 55 | 0 | 55 (100) | 1 per 1000 patch images | 0 (100) | 55 (100) | 100 | 1.0 | 1.0 | 1.0 |
| Gleason Pattern 5 | 55 | 43 (78.1) | 12 (21.8) | 1 per 1000 patch images | 29 (67.4) | 11 (91.6) | 72.7 | 0.967 | 0.674 | 0.795 |
| Cribriform morphology | 55 | 8 (14.5) | 47 (85.5) | 3 per 1000 patch images | 6 (75.0) | 44 (93.6) | 90.9 | 0.75 | 0.667 | 0.706 |
| Ductal morphology | 55 | 44 (80.0) | 11 (20.0) | 1 per 1000 patch images | 43 (97.7) | 8 (72.7) | 92.7 | 0.977 | 0.935 | 0.956 |

We tested the model performance for Gleason pattern 3 (GP3) and pattern 4 (GP4) on an external ISUP dataset (**Figure 1 and Table 3**). Here, the model achieved an AUC of 0.937 and a F1-score of 0.9 for GP3 and 0.83 for GP4. At the case level, GP3 was correctly identified in 97% of cases and all cases with GP4 were detected correctly. The detection model for Gleason pattern 5 (GP5) achieved a F1-score of 0.9 at patch level and TPR of 91.6% at case level.

Table 3: The model accuracies for Gleason pattern detection on an external dataset from ISUP. Each image was evaluated by expert panel members (23 members). This dataset reached a consensus rate of 65% among pathologists. 95% Confidence Interval for uncertainty measurement determined by bootstrapping with 1000 replications)

| Positive set | Negative set | Negative | Positive | AUC (95% CI) | Brier score (95% CI) | Optimal Cut-off of probability | CA | Category | Recall | Precision | F1-score |
|---|---|---|---|---|---|---|---|---|---|---|---|
| Gleason Pattern 3 | Gleason Pattern 4 | 80 | 140 | 0.937 (0.904-0.966) | 0.291 (0.256-0.330) | 0.188 | 0.87 | GP3 | 0.87 | 0.92 | 0.90 |
| | | | | | | | | GP4 | 0.88 | 0.80 | 0.83 |

The cribriform or ductal morphology was detected with AUCs of 0.928 or 0.870 at the patch level. At the case level, TPR for the detection of cribriform morphology was 93.6% with an overall F1-score of 0.706 whereas TPR for the detection of ductal morphology was 72.7% with an overall F1-score of 0.956.

### Real Case Usage of Deep Learning Models for clinician-oriented cMDX reports

**Figure 2A/B** demonstrates the annotation results and illustrates examples of the activation maps for different findings. By visual reviewing the cMDX reports of 55 cases for correctness of the annotation, we found that the tumor lesions were correctly detected and annotated for all cases. However, the prostate cancer detection was irritated by the histology of the ejaculatory ducts and falsely considered a small part of the ejaculatory ducts as tumor area in 4 cases. The accuracy for tumor detection is provided for each case in **Supplement file 1**. The finding list of each lesions were randomly reviewed, and we confirmed that all the finding listed for the lesion were correct.

**Figure 3** provides an overview of the user-interface for viewing the cMDX reports. An example cMDX file and **the viewer tool are provided on GitHub**

**(https://github.com/oeminaga/cmdx_report.git)**. The user interface provides information related to the presence of Prostate Cancer, Gleason patterns 3, 4 and 5, the cribriform and ductal morphology, and the relative/absolute tumor volume. Similar to the original cMDX report editor, the pathologist can provide the tumor grading according to the Gleason grading system [26], the tumor stage using the UICC TNM staging system [27], the extracapsular extension and the surgical margin status. **Supplement file 2** provides an example of cMDX file that includes representative images of the PCa lesions. By Looking at the file sizes, 55 cMDX reports occupied 36.9 gigabytes whereas the corresponding gigapixel histology images required 1.4 terabytes of storage spaces.

*[Removed due to copyright issue]*

*Figure 2A the slide thumbnail (Grid) was used to define a grid to split the histology image into patches, from which the tumor probability was determined for each patch and a heatmap was reconstructed. The generation of the heatmap was repeated for other findings and the heatmap for the prostate cancer was used to determine the lesion boundary (ANNO). PCA: Prostate Cancer; DA: Ductal morphology; CRI: Cribriform morphology; Gleason pattern 3 (GP3), 4 (GP4) or 5. Nerve and/or vessel structures (NERV/VES); inflammation (INF) signatures of high-grade intraepithelial neoplasia (HGPIN); Automated annotation of the lesion (Anno).*

*[Removed due to copyright issue]*

*Figure 2B the activation map for different findings. signatures of high-grade intraepithelial neoplasia (HGPIN)*

*[Removed due to copyright issue]*

*Figure 3: The graphical user interface of the cMDX report viewer (A). The user can access the original gigapixel histology images if these images are available on the local storage. By clicking on the lesion (marked with red color), a second window showing the region appears (B). The user can zoom in or out and scroll through the lesion. The original input data were altered for patient's privacy reasons.*

## Discussion

The current study demonstrated that deep learning models for different histology of prostate pathology are feasible. As real case usage, we demonstrated the feasibility of using these models for annotation tasks of the electronic cMDX (clinical map document) pathology reports as the original framework for the cMDX pathology reports is already part of the clinical routine in Prostate Center of the University Hospital Muenster for more than a decade [24,25]. Our work differs from previous works [11,12,14,28-33] in prostate cancer and Gleason pattern detection in many significant ways. First, the current study covered various types of histology images including the whole-mount slide of a prostatectomy slice, the whole-slide image that contains a portion of the prostatectomy slice, a TMA slide, and internet images (i.e., ISUP images). Second, we utilized histology images that cover all anatomical zones of the prostate and the seminal vesicle. We preferred the whole-slide H&E images of prostatectomy specimens over the biopsy samples for this study for the following reasons. The prostate consists of four major anatomy regions (i.e., peripheral zone, central zone, transition zone and anterior fibromuscular stroma), where the peripheral zone occupies 70% of the prostate [34]. Each prostatic zone has its own histological features that distinguish itself from other zones [34]. Usually, the systematic biopsy scheme targets the peripheral zone as the majority of prostate cancer originates from this zone (68%). However, the remaining prostate cancer is located in other zones which are usually not targeted by the systematic biopsy scheme at the initial biopsy setting due to the low tumor probability in these zones [35,36]. As the maximum cylindric tissue volume for a 18-gauge biopsy side-notch needle with 2.5 cm stroke length is 0.0316 cc [37,38] and the prostate volume ranges between 24 and 106 cc [39,40], a single biopsy core represents a small fraction of the prostate. Thus, the prostate biopsy is not representative due to the heterogeneity of prostate tissue and prostate cancer and is associated with sampling errors [41]. The pathology evaluation of prostatectomy specimens reflects more the real pathological conditions of the prostate cancer and provides more apparent pathological evidences (e.g. tumor heterogeneity, tumor volume and tumor extent) than the pathology evaluation of biopsy cores[42]. Further, the current GG system for biopsy is limited by many challenges associated with sampling errors, high interobserver variations, the biopsy targeting angles and its discrepancy to the final GG in 30-40% cases [4,43]. Thus, the GG system of prostatectomy, if available, is preferred as reference pathology over that of prostate biopsy. For instance, studies evaluating the GG upgrading, a frequent situation in PCa, are considering the GG of the prostatectomy specimens as reference or final GG to identify cases whose final GG are upgraded from the biopsy GG [44,45]. Another example is the Epstein criteria for active surveillance that was developed on the basis of the pathology evaluation of prostatectomy specimens [46,47]. In light of this, the generalizability of the deep learning models that were developed based on biopsy samples or TMA for prostate cancer detection or grading remain questionable. Third, the current study covered different finding families and were evaluated on different datasets for robustness and performance consistency. The cancer detection accuracies remained stable over different types of histology images and scanner types. Our findings show that our model based on the PlexusNet architecture[23] performed well in prostate cancer detection and Gleason patterns although it was developed by using 12.7% of the total histology images. The discrepancy in the performance of the HGPIN detection model between the internal validation set annotated by YT and the external validation set annotated by CK is due to the inconsistency in the definition of HGPIN lesion as the current inter-observer agreement for HPGIN is 70% according to Iczkowski et al [48]. Fourth, we provided a real case usage of our models by integrating the detection models into the electronic cMDX pathology report. The cMDX pathology report was designed from the urological aspect and includes information relevant for tumor classification (pT) from whole-mounted prostatectomy specimens such as the tumor spatial distributions[27] and the presence of Gleason patterns. Using the detection model for prostate cancer facilitated a very accurate estimation of the tumor volume related to the ground truth ($R^2$:0.987). One of the challenges of reporting tumor volume is the accurate estimation as it is one of the reasons for controversy in the predictive value of tumor volume or relative tumor volume [49-53]. Although no consensus method has emerged for measuring tumor volume, ISUP advocates for technological advances in imaging techniques to reinforce the clinical rationale for incorporating a size-related staging parameter into the pathological reporting of prostate cancers [54]. Finally, we followed the clinical guidelines for pathological evaluation and considered the needs of urologists for histopathological information in the cMDX reports [24]. Fifty-five cases were tested on a single GPU and the models were also trained on a single GPU (Titan V with 11 GB VRAM) and 2 TB PCIe flash memory, where one case required 35+/-6 minutes in average to complete the all finding detections. This duration includes the time cost for input/output access that has impacts on the processing speed. Thus, our models are energy efficient compared to models that require multiple GPUs or expensive GPU cloud solutions for training. We believe that energy-/cost-effective AI-based solutions will receive more acceptance in healthcare as a recent survey showed that the majority of U.S. public (69%) advocates the need for prioritizing the reduction of healthcare cost by the U.S. government [55].

Deep learning has now facilitated the automation of the time-consuming annotation procedure for the tumor extent and helped to shorten the considerable documentation duration required for the manual delineation of the tumor extent [25]. Given that there is no standard validation set for prostate cancer to compare with results from other studies, we explicitly avoided any comparison with previous studies. In our opinion, performing a model comparison is artificial and challenging because the model optimization depends on the developers and the data preparation. Additionally, there is no standard configuration for hyperparameters or augmentations for the existing models for prostate cancer and related findings. The condition of the input data (e.g. the magnification level and patch size), configurations of hyperparameters (e.g. batch size) or augmentations (e.g. the degree of rotation) lead to different performance despite having the same model architecture [14,23,56,57]. Moreover, there are so many deep learning architecture and many trimmed versions of "state-of-art" models making a reasonable model comparison difficult for one research team to cover the all existing deep learning models [14,28,33,58-60]. Therefore, we advocate providing detailed information related to the model architecture, the hardware, the hyperparameter and augmentation configuration, and conducting the model evaluation on a standard validation set to achieve a reasonable model comparison. To support the standardized performance reporting for prostate cancer detection and related finding, the image file list, the annotation data for TCGA images will be available for non-commercial research.

The current study inherits some limitations that warrant mention. First, this study has a retrospective character and therefore encompasses the limitations of a retrospective study. Although we implemented a quality control procedure for the blurriness and brightness of histology images, the protective measurement may have failed to identify poor-quality images that may have impacted the model performance. Thus, a periodic quality control of histology images should be made prior to feeding the framework with histology images. There is a need to adapt the deployment of DL models to the existing infrastructure and resources. Further, the definition of the thresholds for cancer detection varies according to the application as the evaluation conditions for biopsy cores differs from for the evaluation conditions of prostatectomy specimens. Other limitations include the high expense and maintenance costs of the infrastructure to digitalize histology slides that continue to restrict the wide-spread usage of digital pathology. We believe that this issue can be resolved by having more competitors in this field to lower the costs to benefit small and midsize healthcare services. A potential limitation of this study is that many pathologists were involved in the annotation procedures. However, such conditions actually mimic the clinical routine and the classification performance for PCA, GP3 and GP4 detection were comparable between different datasets with histology images annotated by different pathologists. We didn't perform any comparison to the human readers as such comparisons are artificial and doesn't represent the clinical routine; The clinical routine includes a close communication between different clinical disciplines and physicians through many channels (e.g. hospital information systems, tumor boards, consulting etc..) and it is well-known that prior knowledge about the clinical information enhances the pathology evaluation [9,10]. Another limitation is that the classification accuracy for Gleason pattern 5 (GP5) was moderate due to the low number and size of the lesions with GP5 as Gleason pattern 5 of our cases are tertiary Gleason patterns and the patients with GG 9-10 are often not amenable to surgical intervention and instead receiving hormonal deprivation and radiation therapy [61,62]. However, the detection model for GP5 has a space for accuracy improvement and we aim to recruit more cases with Gleason pattern 5 to enhance the accuracy of the GP5 detection model over the time. Finally, we focused only on major findings related to prostate pathology and didn't consider all aspects of prostate pathology. However, the purpose of the current study is to show that DL is feasible to determine different morphologies of prostate pathology and we do plan to expand the coverage to benign hyperplasia and intraductal prostate cancer based on the existing highly curated datasets. Our future work will focus on integrating the DL into the cMDX framework and conduct research evaluating the benefits of applying DL trained on prostatectomy specimens for biopsy pathology.

## Conclusion

The current study introduces deep learning models for different histology of prostate pathology deployable for cMDX pathology report generator; it has high accuracies for cancer detection and the detection of related findings.

## Material and Methods

This study used prospectively collected whole-slide diagnostic histology images (TCGA-PRAD) from TCGA (The Cancer Genome Atlas) and Stanford University in accordance with the privacy regulations and the Helsinki declaration. The study was approved by the IRB (IRB-46418). The histology images were stained with Hematoxylin and Eosin staining (H&E) and acquired using an Aperio Digital Pathology Slide Scanner -Scanner type A- from Leica Biosystem (Wetzlar, Germany). The TCGA images were scanned at a 40x objective zoom, whereas the Stanford images were scanned at a 20x objective zoom. These images were stored in SVS format. Our cohort consisted of 449 H&E images from TCGA; 466 whole-mount H&E images from 65 cases who underwent radical prostatectomy were also considered. Additionally, we included 125 whole-slide images representing the index lesions in 125 cases from the historic McNeal dataset that were scanned at 40x objective zooming level using a slide scanner from Philips (Amsterdam, Netherland) – Scanner B. A tissue micro array (TMA) from 339 prostatectomy specimens with 932 spots from prostate cancer index lesions and 197 spots with normal tissues was stained with H&E and scanned using the Ariol microscope system manufactured by Leica -Scanner C- (Wetzlar, Germany). Forty-two spot images from a second TMA that have, in addition to normal tissue and prostate cancer, prostatic intraepithelial neoplasia was also included in our study. Finally, 220 H&E histology images from the International Society of Urological Pathology (ISUP) reference library images were included (Internet, Unknown scanner vendor). In total, we collected 2,431 H&E images that spanned a variety of image qualities, origins (WS, TMA, WM, Internet), scanning machines, timestamps, H&E staining protocols, and institutions. All histology images were capture using the bright field microscopy. For real case usage, we applied the existing cMDX framework for generating pathology reports for prostatectomy specimens and included the automated annotation of prostate cancer and related finding. A detailed description of cMDX framework can be obtained from Eminaga et al [24].

## Cohort for Prostate Cancer Detection

The development set was randomly selected and consisted of 250 histology images from TCGA (55% of TCGA images) and 60 whole-mount histology images from 10 Stanford cases (12% of Stanford WM images). The main reason of including 10 cases from Stanford is the high tumor burden of TCGA images (Mean pixel number with tumor in percentage: 45+/-4%), which does not cover all aspects of histological structures of the prostate (e.g., ejaculatory duct, different forms of the benign hyperplasia, epithelial tissues from central zones, urethra). Therefore, we selected these images from Stanford that have tumor burdens below 10% of the prostate and exhibit different prostatic anatomic structures. The development set was then randomly split into a training set (80%), and a test set for internal validation (20%). The validation set for model training was generated by randomly selecting 10% of patch images from each case of the test set.

All images were annotated for tumor lesions by experienced board-certified pathologists (CK, YT, MA and RW) and a urologist (OE) who all have significant experience in research related to the pathology of prostate cancer and its associated findings. The prostate cancer lesions of the whole-mount (WM) images were annotated by CK. The annotation of the tumor lesions on the regular whole-slide (WS) images from TCGA-PRAD and McNeal's dataset was made by OE and confirmed by MA for the correctness of the annotation. The tissue micro arrays (TMA) for prostate cancer was already created according to the tumor status. The spots with tumors were obtained from lesions in prostatectomy specimens that were identified by many pathologists during the clinical routine. After creating the TMA, the tumor status of each TMA spot was evaluated and confirmed by RW.

For external validation, we utilized three datasets with different data origins. The first validation set consisted of 254 whole-mount H&E images from serially sectioned prostatectomy specimens. The second validation set had 13 whole-slice H&E images from the McNeal dataset. The third external data set with H&E images came from the Stanford Tissue Microarray (TMA) Database with prostate cancer (n= 1,129) and was applied for evaluating detection performance. These histology spot images (Size: 1,024x1,024 pixels) were stained with H&E, captured at 20x objective magnification level. We clipped the middle region of the spot image which contains the tissue with relevant findings by 512x512 pixels and applied the repeated fill effect with the clipped image for a new image with a size of 1,024x1,024 pixels, which was then resized to 512x512 pixels for each H&E image.

To evaluate the inter-observer annotation agreement, 6 whole-mount images from Stanford were independently annotated for the prostate cancer lesions by YT (OE refined the marked lesion boundaries) and KC and the inter-observer agreement was estimated by Cohen Kappa after setting a grid with tiles of 512x512 pixels for each image.

## Cohort for Findings related to Prostate Cancer

Sixty-four H&E whole slice images were randomly selected from the development set including 58 images from Stanford and 6 images from TCGA dataset. YT annotated the regions of interests (ROI) covering all findings listed in **Supplement file 3** and the annotation contours of ROI were refined by OE. Gleason grading was made in accordance with ISUP guidelines from 2016 [63]. The regions of interest were tiled by 512x512 pixels, and the resulting patch images were split after the stratification by case into the development sets (70%, n=44) and internal validation sets (30%, n=20). From the development set, we generated a training set with 90% of the development set and the remaining lesions were assigned to the model validation set. In order to account for class imbalance (arbitrary defined by a ratio of 8:1 for the majority and minority classes) when it occurred, we applied the oversampling of the minority class to increase the frequency of the patch images from the minority class. Supplement file 3 provides information regarding the findings that have the imbalance class problems and the applied factor to oversample the minority class for solving this problem.

It is worth noting that the internal validation set (test set) consists cases that have a detailed morphology annotation as given in **Supplement file 1 and 3**. Further, we visually checked subsets of patch images from the internal validation set for the presence of Gleason patterns 3, 4, and 5 to ensure that patch images represent the corresponding findings (**Supplement file 1**). We didn't consider the Gleason patterns less than 3 in our study as these patterns are no longer utilized in clinical routine due to the lack of the clinical implication[4].

Gleason grading plays an important role in clinical decision making and we wanted to ensure that our models for Gleason pattern 3 and 4 provide results comparable to those of experts. Therefore, we utilized the International Society of Urological Pathology (ISUP) reference library images for Gleason grading, which were graded by a majority voting of a panel of 23 expert members of ISUP, to externally validate our models for Gleason pattern 3 and 4 [64]. Here, we considered 220 H&E images (Size: 2048x2048 pixels, captured at 20x magnification level) having either 4+4 or 3+3 Gleason score for our evaluation to limit the risk of the inaccurate evaluation and finding uncertainty in each patch image and to increase the likelihood of the presence of a single finding in each patch image. Further, we wanted to ensure that our models can correctly detect the Gleason patterns from ISUP histology images as these images are considered as Gold standard and used for education purposes. We were unable to evaluate Gleason pattern 5, given that there are 6 images with Gleason pattern 5 concurrent with Gleason pattern 4. High-grade Intraprostatic intraepithelial neoplasia (HGPIN), the most presumed precursor of prostate cancer [65,66], is optionally reported in the pathology reports [67]. So, it was important to validate the model for HGPIN detection by using an external validation set from a TMA containing HGPIN (20 of 42 spots). Spots with HGPIN were labelled by a single experienced uro-pathologists (CK), whereas the development set with HGPIN lesions from WM images was annotated by YT and the marked boundaries of the lesions were refined by OE.

Since the validation sets were acquired using scanners other than those used for the development set, we optimized the brightness of the patch images by multiplying with a scanner factor that may range between 0.01 and 1. The determination of the scanner factor is based on the brute force approach, which finds the best scanner factor by determining the best ROC performance of the model on 5 positive and 5 negative patch images from the new dataset with no need to re-train the model for new slide images captured by scanners other than we used for the development set. The screening for the best scanner factor was made at two steps by the sequential increasing of the scanner factor initially by 0.1 and then by 0.01 in a closed range containing the best factor from the initial screening.

## Generation and Labelling of Patch Images

To define the coordinate grid for patch image generation, the smallest level of the SVS whole-slide image was converted to grayscale. Then, the tissue region was masked by thresholding at the mean gray value of the gray intensity. To determine the coordinates of each patch image, the default patch size (512x512 pixels) was rescaled after dividing by scale factors for height and width. These scale factors were determined by calculating the ratio of the dimension of the whole-slide image at 10x to the image dimension of the highest level. Patches were generated with an overlap ratio of 0.2. Patches not covered by the masked tissue region were excluded in order to remove background images from the dataset. Finally, the grid for the patch images was upscaled after multiplying by the scale factors. All histological images were tiled by 512x512 pixels (330x330 µm) at a 10x magnification level based on the grid coordinates. For labelling patch images used for training and validation, the ground truth was considered as a binary mask generated based on the annotation data that covered all findings relevant for the current study in each slide image. We developed a custom patch image generator that generates these masks for model training and evaluation by conducting a key search for required findings in the annotation data. Each finding was annotated on the slide images independently from other findings. The definition of the negative set depends on the target finding as given in **Supplement** file 1. The patch mask is extracted at the same location of the corresponding patch image. The percentage of positive pixels to the total image size was estimated to label each patch image according to the binary classification. A patch image is positive if the number of positive pixels meets or exceeds a threshold of 20%. By determining the threshold, the effect of potential errors associated with the annotation procedure was taken into account, especially in the edges of the annotated areas as the edge areas are more prone to annotation errors and false-positive conditions than other parts of the annotated region. We also estimated the threshold by building an analogy to the risk of prostate cancer ranges between 15-25% by a threshold of 3-4 ng/mL for the serum level of prostate-specific antigen, where urologists usually consider an active measurement by the given risk for prostate cancer [68].

## Color Intensity Optimization for Hematoxylin Eosin for aged whole-slide H&E images

A long storage period of H&E slides from McNeal datasets > 10 years and aging processes caused paled H&E staining of these slides. Specially, the nuclei staining is affected the most by aging. In order to reconstruct the color intensity, we developed an algorithm specific for the color intensity correction of H&E McNeal images inspired by Macenko's approach [69]. Before feeding the patch images for any prediction procedures, we converted the RGB color space of the patch images into the optical density (OD) space for red, green and blue channels. Then, we restricted the OD ranges between 0.5 and 0.95 and excluded extreme values of OD. After that, we calculated the covariance matrix of a single patch image first by combining the color channels with itself and then with each other and calculated the mean covariance matrix from all channel combinations. Finally, the eigen vector was calculated using the mean covariance matrix and equalized to the stain vector. The determination of the stain vector occurs once for each WS H&E image using the first patch image.

The obtained stain vector is applied to optimize the color intensity of nuclei for all patch images originated from McNeal's whole-slide H&E images by multiplying the stain vector with the OD for each patch image. Finally, the OD matrices for red and green and blue channels were converted back to the matrices with the RGB color space. The patch image is corrected by merging the converted matrices and the original RGB patch image. An example showing a patch image before and after applying the H&E color optimization is provided in **Supplement file 1**.

## Deep Learning models for Cancer Detection and Detection of other findings relevant for pathology report

**Supplement file 3** provides information related to the cohort constitution, the model architecture, and the hyperparameters applied for each finding considered in our screening report. Additional information about the CNN architecture of each model can also be obtained from **supplement file 1**. Most models were trained using the optimization algorithm Adaptive Moment Estimation (i.e., ADAM) instead of Stochastic Gradient Descent [70]. The gradient Noise was applied for models to improve the model learning as the noise induced by the stochastic process aids generalization by reducing overfitting [71]. The maximum number of training epochs was set to either to 50 or 20 and an early stopping algorithm was used to stop training after five consecutive epochs with not improvement in the classification accuracy. The batch size was defined as 16. Relevant findings other than cancer were delineated on the randomly selected 62 H&E images from Stanford and TCGA by OE under supervision of MA and YT. The relevant findings and their proportions in the development and test sets for each model are listed in **Supplement file 1 and 3**. After that, the regions of interests were extracted and tiled by 512x512 pixels and applying an overlap rate of 0.5. These tiled images were split into a training set, validation set, and test set. CLAHE (Contrast Limited Adaptive histogram equalization) was applied for some models in order to optimize the image contrast and all pixel values were normalized by 1/255. We applied class weighting, oversampling, and image augmentation of rare findings to reduce the class imbalance effect and increase the presence probability of these scarce findings in the batch sequence during the model training.

The image augmentation included rotation, horizontal and vertical flips, image shearing, and zooming and brightness manipulation. Additionally, we applied random RGB channel shifting and random modification of the image quality by changing the JPEG compression rate for certain findings.

## Planimetric Cancer volume estimation

The prostate volume was calculated after formalin fixation by weighing the prostate specimen without the seminal vesicles. For the purpose of our study, the prostate weight in grams was considered roughly equivalent to its volume in cubic centimeters ($cm^3$); the tumor/entire gland ratio is then used to calculate the volume of the tumor in $cm^3$. A correction factor for tissue shrinkage after formalin fixation was not considered. The computational tumor volume estimate was performed on the basis of the volumetric calculation. Every tumor focus in each slice is estimated by counting the pixels affected by PCa. The cancer area is then divided by the slice area occupied by the prostatic slice and then added to calculate the relative cancer volume. Finally, the total relative cancer volume is multiplied with the prostate volume to calculate the cancer volume in cm$^3$.

## Study Cohort for the accuracy evaluation of pathology screening reports

Slides of sequential whole-mount slices from 55 cases that underwent prostatectomy were scanned at a 20x objective zoom and then fed into the cMDX framework. The whole-mount H&E histology slides were digitalized for all slices of the prostate from prostatectomy specimens in each case. The Gleason patterns were extracted from pathology reports of these cases using a natural language process and the keyword search. These extracted findings were further checked for correctness by manually reviewing the pathology reports and then compared with the reported Gleason pattern in the automated pathology reports. The pathology reports and H&E slide images were evaluated for the presence of ductal and cribriform morphology at the case level ( OE evaluated the pathology reports and shared anonymized histology images with MA for evaluation). The relative tumor volume was compared between the ground truth annotation made by CK and the relative tumor volume calculated by the cMDX pathology report.

## Evaluation metrics

The classification performance of the final test set for pathological findings was evaluated once using classification accuracy, precision, recall, F-measure (F1 score), Area Under of the Receiver Operating Characteristics curve (AUC-ROC), and Brier score. F1-score is the harmonic mean between of the precision and recall applied for the measurement of the classification performance and imbalanced classification problems. The Fowlkes-Mallows index (FMI) is defined as the geometric mean between of the precision and recall and generally used for the similarity measurement between two groups. The Brier score measures the accuracy of probabilistic predictions for binary outcome and can also be used as "calibration" measurement of the prediction model. We evaluated the classification performance slide-wise, spot-wise, and patch-wise for prostate cancer, and case-wise, patch-wise or spot-wise for Gleason patterns 3 and 4. The presence of Gleason pattern 5 was evaluated at the case level. The classification performance of the framework for ductal and cribriform morphology was evaluated at the case level as well. Given that vessel and nerves are widely spread inside the prostate, we considered only the internal validation. The coefficient of the regression score determined the correlation of relative tumor volumes between the ground truth and the cMDX framework at case level. The pair-wise student t-test was applied to identify the significance of variation between the ground truth and the cMDX/PlexusNet-based tumor extent detection for relative tumor volume. The reported p-value is two-sided and statistical significance was assumed as $P ≤ 0.05$.

Our analyses were based on Python 3.6 (Python Software Foundation, Wilmington, DE) or R 3.5.1 (R Foundation for Statistical Computing, Vienna, Austria) and applied the Keras library which is built-on the TensorFlow framework, to develop the models. All analyses were performed on a GPU machine with 32-core AMD processor with 128 GB RAM (Advanced Micro Devices, Santa Clara, CA), 2 TB PCIe flash memory, 5 TB SDD Hard disks, and a single NVIDIA Titan V GPU with 12 GB VRAM.

## Acknowledgment
PlexusNet and the derivate digital markers for survival and genomic alteration are patented.

## Reference

1. Jemal, A., Siegel, R., Xu, J. & Ward, E. Cancer statistics, 2010. *CA Cancer J Clin* **60**, 277-300 (2010).
2. Abdollah, F., *et al.* A competing-risks analysis of survival after alternative treatment modalities for prostate cancer patients: 1988-2006. *European urology* **59**, 88-95 (2011).

3. Epstein, J.I., Srigley, J., Grignon, D. & Humphrey, P. Recommendations for the reporting of prostate carcinoma: Association of Directors of Anatomic and Surgical Pathology. *American journal of clinical pathology* **129**, 24-30 (2008).
4. Egevad, L., Delahunt, B., Srigley, J.R. & Samaratunga, H. International Society of Urological Pathology (ISUP) grading of prostate cancer - An ISUP consensus on contemporary grading. *APMIS* **124**, 433-435 (2016).
5. Fine, S.W.*, et al.* A contemporary update on pathology reporting for prostate cancer: biopsy and radical prostatectomy specimens. *Eur Urol* **62**, 20-39 (2012).
6. Elmore, J.G.*, et al.* Pathologists' diagnosis of invasive melanoma and melanocytic proliferations: observer accuracy and reproducibility study. *bmj* **357**, j2813 (2017).
7. Shen, S. & Ma, W.J. Variable precision in visual perception. *Psychological review* **126**, 89 (2019).
8. Madden, D.J. Adult age differences in the attentional capacity demands of visual search. *Cognitive Development* **1**, 335-363 (1986).
9. Martins, T.R.*, et al.* Influence of Prior Knowledge of Human Papillomavirus Status on the Performance of Cytology Screening. *Am J Clin Pathol* **149**, 316-323 (2018).
10. Nakhleh, R.E., Gephardt, G. & Zarbo, R.J. Necessity of clinical information in surgical pathology. *Arch Pathol Lab Med* **123**, 615-619 (1999).
11. Arvaniti, E.*, et al.* Automated Gleason grading of prostate cancer tissue microarrays via deep learning. *Sci Rep* **8**, 12054 (2018).
12. Arvaniti, E.*, et al.* Author Correction: Automated Gleason grading of prostate cancer tissue microarrays via deep learning. *Sci Rep* **9**, 7668 (2019).
13. Li, J.*, et al.* An EM-based semi-supervised deep learning approach for semantic segmentation of histopathological images from radical prostatectomies. *Comput Med Imaging Graph* **69**, 125-133 (2018).
14. Campanella, G.*, et al.* Clinical-grade computational pathology using weakly supervised deep learning on whole slide images. *Nature Medicine* (2019).
15. Nagpal, K.*, et al.* Development and validation of a deep learning algorithm for improving Gleason scoring of prostate cancer. *npj Digital Medicine* **2**, 48 (2019).
16. Li, J.*, et al.* An attention-based multi-resolution model for prostate whole slide imageclassification and localization. *arXiv preprint arXiv:1905.13208* (2019).
17. Lawson, P., Schupbach, J., Fasy, B.T. & Sheppard, J.W. Persistent homology for the automatic classification of prostate cancer aggressiveness in histopathology images. in *Medical Imaging 2019: Digital Pathology*, Vol. 10956 109560G (International Society for Optics and Photonics, 2019).
18. He, K., Zhang, X., Ren, S. & Sun, J. Deep residual learning for image recognition. in *Proceedings of the IEEE conference on computer vision and pattern recognition* 770-778 (2016).
19. Simonyan, K. & Zisserman, A. Very deep convolutional networks for large-scale image recognition. *arXiv preprint arXiv:1409.1556* (2014).
20. Szegedy, C., Vanhoucke, V., Ioffe, S., Shlens, J. & Wojna, Z. Rethinking the inception architecture for computer vision. in *Proceedings of the IEEE conference on computer vision and pattern recognition* 2818-2826 (2016).

21. Iandola, F., *et al.* Densenet: Implementing efficient convnet descriptor pyramids. *arXiv preprint arXiv:1404.1869* (2014).
22. Yosinski, J., Clune, J., Bengio, Y. & Lipson, H. How transferable are features in deep neural networks? in *Advances in neural information processing systems* 3320-3328 (2014).
23. Eminaga, O.*, et al.* Plexus Convolutional Neural Network (PlexusNet): A novel neural network architecture for histologic image analysis. *arXiv preprint arXiv:1908.09067* (2019).
24. Eminaga, O.*, et al.* CMDX(c)-based single source information system for simplified quality management and clinical research in prostate cancer. *BMC Med Inform Decis Mak* **12**, 141 (2012).
25. Eminaga, O.*, et al.* Clinical map document based on XML (cMDX): document architecture with mapping feature for reporting and analysing prostate cancer in radical prostatectomy specimens. *BMC Med Inform Decis Mak* **10**, 71 (2010).
26. Gleason, D.F. & Mellinger, G.T. Prediction of prognosis for prostatic adenocarcinoma by combined histological grading and clinical staging. *J Urol* **111**, 58-64 (1974).
27. Brierley, J.D., Gospodarowicz, M.K. & Wittekind, C. *TNM Classification of Malignant Tumours*, (Wiley, 2016).
28. Lucas, M.*, et al.* Deep learning for automatic Gleason pattern classification for grade group determination of prostate biopsies. *Virchows Arch* (2019).
29. Achi, H.E.*, et al.* Automated Diagnosis of Lymphoma with Digital Pathology Images Using Deep Learning. *Ann Clin Lab Sci* **49**, 153-160 (2019).
30. Fischer, W., Moudgalya, S.S., Cohn, J.D., Nguyen, N.T.T. & Kenyon, G.T. Sparse coding of pathology slides compared to transfer learning with deep neural networks. *BMC Bioinformatics* **19**, 489 (2018).
31. Jiang, Y., Chen, L., Zhang, H. & Xiao, X. Breast cancer histopathological image classification using convolutional neural networks with small SE-ResNet module. *PLoS One* **14**, e0214587 (2019).
32. Nagpal, K.*, et al.* Development and validation of a deep learning algorithm for improving Gleason scoring of prostate cancer. *NPJ Digit Med* **2**, 48 (2019).
33. Litjens, G.*, et al.* Deep learning as a tool for increased accuracy and efficiency of histopathological diagnosis. *Sci Rep* **6**, 26286 (2016).
34. McNeal, J.E. The zonal anatomy of the prostate. *Prostate* **2**, 35-49 (1981).
35. McNeal, J.E., Redwine, E.A., Freiha, F.S. & Stamey, T.A. Zonal distribution of prostatic adenocarcinoma. Correlation with histologic pattern and direction of spread. *Am J Surg Pathol* **12**, 897-906 (1988).
36. Rouviere, O.*, et al.* Use of prostate systematic and targeted biopsy on the basis of multiparametric MRI in biopsy-naive patients (MRI-FIRST): a prospective, multicentre, paired diagnostic study. *Lancet Oncol* **20**, 100-109 (2019).
37. Bostwick, D.G. Evaluating prostate needle biopsy: therapeutic and prognostic importance. *CA Cancer J Clin* **47**, 297-319 (1997).
38. Kanao, K.*, et al.* Impact of a novel biopsy instrument with a 25-mm side-notch needle on the detection of prostate cancer in transrectal biopsy. *Int J Urol* **25**, 746-751 (2018).

39. Berges, R. & Oelke, M. Age-stratified normal values for prostate volume, PSA, maximum urinary flow rate, IPSS, and other LUTS/BPH indicators in the German male community-dwelling population aged 50 years or older. *World J Urol* **29**, 171-178 (2011).
40. Pinsky, P.F.*, et al.* Prostate volume and prostate-specific antigen levels in men enrolled in a large screening trial. *Urology* **68**, 352-356 (2006).
41. Bjurlin, M.A. & Taneja, S.S. Standards for prostate biopsy. *Curr Opin Urol* **24**, 155-161 (2014).
42. Elgamal, A.A.*, et al.* Impalpable invisible stage T1c prostate cancer: characteristics and clinical relevance in 100 radical prostatectomy specimens--a different view. *J Urol* **157**, 244-250 (1997).
43. Epstein, J.I., Feng, Z., Trock, B.J. & Pierorazio, P.M. Upgrading and downgrading of prostate cancer from biopsy to radical prostatectomy: incidence and predictive factors using the modified Gleason grading system and factoring in tertiary grades. *Eur Urol* **61**, 1019-1024 (2012).
44. De Nunzio, C.*, et al.* The new Epstein gleason score classification significantly reduces upgrading in prostate cancer patients. *Eur J Surg Oncol* **44**, 835-839 (2018).
45. Epstein, J.I.*, et al.* A Contemporary Prostate Cancer Grading System: A Validated Alternative to the Gleason Score. *Eur Urol* **69**, 428-435 (2016).
46. Ross, H.M.*, et al.* Do adenocarcinomas of the prostate with Gleason score (GS) </=6 have the potential to metastasize to lymph nodes? *Am J Surg Pathol* **36**, 1346-1352 (2012).
47. Epstein, J.I., Walsh, P.C., Carmichael, M. & Brendler, C.B. Pathologic and clinical findings to predict tumor extent of nonpalpable (stage T1c) prostate cancer. *JAMA* **271**, 368-374 (1994).
48. Iczkowski, K.A.*, et al.* Intraductal carcinoma of the prostate: interobserver reproducibility survey of 39 urologic pathologists. *Ann Diagn Pathol* **18**, 333-342 (2014).
49. Wolters, T.*, et al.* Should pathologists routinely report prostate tumour volume? The prognostic value of tumour volume in prostate cancer. *Eur Urol* **57**, 821-829 (2010).
50. Hong, M.K.*, et al.* Prostate tumour volume is an independent predictor of early biochemical recurrence in a high risk radical prostatectomy subgroup. *Pathology* **43**, 138-142 (2011).
51. Carvalhal, G.F.*, et al.* Visual estimate of the percentage of carcinoma is an independent predictor of prostate carcinoma recurrence after radical prostatectomy. *Cancer* **89**, 1308-1314 (2000).
52. Perera, M., Lawrentschuk, N., Bolton, D. & Clouston, D. Comparison of contemporary methods for estimating prostate tumour volume in pathological specimens. *BJU Int* **113 Suppl 2**, 29-34 (2014).
53. Hinkelammert, R.*, et al.* Tumor percentage but not number of tumor foci predicts disease-free survival after radical prostatectomy especially in high-risk patients. *Urol Oncol* **32**, 403-412 (2014).
54. van der Kwast, T.H.*, et al.* International Society of Urological Pathology (ISUP) Consensus Conference on Handling and Staging of Radical Prostatectomy Specimens. Working group 2: T2 substaging and prostate cancer volume. *Mod Pathol* **24**, 16-25 (2011).

55. Blendon, R.J., Benson, J.M. & McMurtry, C.L. The Upcoming U.S. Health Care Cost Debate - The Public's Views. *N Engl J Med* **380**, 2487-2492 (2019).
56. Bergstra, J., Yamins, D. & Cox, D.D. Making a science of model search: Hyperparameter optimization in hundreds of dimensions for vision architectures. (2013).
57. Snoek, J., Larochelle, H. & Adams, R.P. Practical bayesian optimization of machine learning algorithms. in *Advances in neural information processing systems* 2951-2959 (2012).
58. Tan, M. & Le, Q.V. EfficientNet: Rethinking Model Scaling for Convolutional Neural Networks. *arXiv preprint arXiv:1905.11946* (2019).
59. Alom, M.Z.*, et al.* A state-of-the-art survey on deep learning theory and architectures. *Electronics* **8**, 292 (2019).
60. Wei, J.W.*, et al.* Pathologist-level classification of histologic patterns on resected lung adenocarcinoma slides with deep neural networks. *Sci Rep* **9**, 3358 (2019).
61. Al-Hussain, T.O., Nagar, M.S. & Epstein, J.I. Gleason pattern 5 is frequently underdiagnosed on prostate needle-core biopsy. *Urology* **79**, 178-181 (2012).
62. Mohler, J.L.*, et al.* Prostate Cancer, Version 2.2019, NCCN Clinical Practice Guidelines in Oncology. **17**, 479 (2019).
63. Epstein, J.I. International Society of Urological Pathology (ISUP) Grading of Prostate Cancer: Author's Reply. *Am J Surg Pathol* **40**, 862-864 (2016).
64. Egevad, L.*, et al.* Pathology Imagebase-a reference image database for standardization of pathology. *Histopathology* **71**, 677-685 (2017).
65. Bostwick, D.G. & Cheng, L. Precursors of prostate cancer. *Histopathology* **60**, 4-27 (2012).
66. Jung, S.H.*, et al.* Genetic Progression of High Grade Prostatic Intraepithelial Neoplasia to Prostate Cancer. *Eur Urol* **69**, 823-830 (2016).
67. Humphrey, P.A., Moch, H., Cubilla, A.L., Ulbright, T.M. & Reuter, V.E. The 2016 WHO Classification of Tumours of the Urinary System and Male Genital Organs-Part B: Prostate and Bladder Tumours. *Eur Urol* **70**, 106-119 (2016).
68. Vickers, A.J.*, et al.* The relationship between prostate-specific antigen and prostate cancer risk: the Prostate Biopsy Collaborative Group. *Clin Cancer Res* **16**, 4374-4381 (2010).
69. Macenko, M.*, et al.* A method for normalizing histology slides for quantitative analysis. in *2009 IEEE International Symposium on Biomedical Imaging: From Nano to Macro* 1107-1110 (IEEE, 2009).
70. Kingma, D.P. & Ba, J. Adam: A method for stochastic optimization. *arXiv preprint arXiv:1412.6980* (2014).
71. Neelakantan, A.*, et al.* Adding gradient noise improves learning for very deep networks. *arXiv preprint arXiv:1511.06807* (2015).